\documentclass[10pt]{article}
\usepackage{a4wide}
\usepackage{amsfonts}
\usepackage[dvipdfm]{graphicx}
\usepackage{amssymb,amsmath}
\usepackage[all]{xy}
\usepackage{pstricks, pst-node}
\usepackage{multicol}   

\usepackage[round,numbers,sort&compress]{natbib}

\newcommand{\kA}{\ensuremath{k_A}}
\newcommand{\kB}{\ensuremath{k_B}}
\newcommand{\kM}{\ensuremath{k_M}}
\newcommand{\kR}{\ensuremath{k_R}}

\newcommand{\eucerr}{\ensuremath{\epsilon_E}}

\newcommand{\abasin}{\ensuremath{A^\#}}
\newcommand{\bbasin}{\ensuremath{B^\#}}
\newcommand{\E}{\ensuremath{\mathbb{E}}}
\newcommand{\Em}{\ensuremath{\mathbb{E}^{-1}}}
\newcommand{\Ep}[1]{\ensuremath{\mathbb{E}_{#1}}}
\newcommand{\Epm}[1]{\ensuremath{\mathbb{E}^{-1}_{#1}}}

\newcommand{\charlierlong}[3]{\ensuremath{C_{#1}\big(#3; #2 \big)}}

\date{}

\begin{document}

\begin{titlepage}

\begin{center}
{\LARGE Perfect Sampling of the Master Equation for Gene
Regulatory Networks}

\vspace{.25in}

\large{Martin Hemberg and Mauricio Barahona\footnote{
    Corresponding author.
Address: Department of Bioengineering, Imperial College London,
South Kensington Campus, London, SW7 2AZ (United Kingdom);
Telephone: +44 207 594 5189; Fax: +44 207 594 6897;
e-mail: m.barahona@imperial.ac.uk} \\
  Department of Bioengineering, \\
  Imperial College London, \\ United Kingdom }

\end{center}

\vspace{.25in}

\abstract{We present a Perfect Sampling algorithm that can be
applied to the Master Equation of Gene Regulatory Networks (GRNs).
The method recasts Gillespie's Stochastic Simulation Algorithm
(SSA) in the light of Markov Chain Monte Carlo methods and
combines it with the Dominated Coupling From The Past (DCFTP)
algorithm to provide guaranteed sampling from the stationary
distribution.  We show how the DCFTP-SSA can be generically
applied to genetic networks with feedback formed by the
interconnection of linear enzymatic reactions and nonlinear Monod-
and Hill-type elements. We establish rigorous bounds on the error
and convergence of the DCFTP-SSA, as compared to the standard SSA,
through a set of increasingly complex examples. Once the building
blocks for GRNs have been introduced, the algorithm is applied to
study properly averaged dynamic properties of two experimentally
relevant genetic networks: the toggle switch, a two-dimensional
bistable system, and the repressilator, a six-dimensional genetic
oscillator.}


\end{titlepage}

\clearpage


\section*{Introduction}

The recent interest in stochastic models of chemical reactions has
been largely motivated by experiments that have demonstrated the
stochastic nature of key processes in the cell, such as signalling
or gene regulation~\citep{Acar:2005, Austin:2006, Blake:2003,
Elowitz:2000, Gardner:2000, Pedraza:2005, Volfson:2006, Cai:2006,
Fung:2005, Becskei:2001}. The inherent stochasticity of these
biochemical processes is due to the low number of molecules
involved in the reactions~\citep{McAdams:1997}, amplified in some
cases by the proximity to a critical point of the dynamics of the
system~\citep{Elf:2003b}. Low copy numbers lead to intrinsic
fluctuations and to the breakdown of models based on differential
equations~\citep{vanKampen:1992}. A number of stochastic models of
gene regulation \citep{Hasty:2000, Kaern:2005, Kepler:2001,
McAdams:1997, Paulsson:2004, Thattai:2001, Bratsun:2005,
Erban:2006, Swain:2002} and signalling \citep{Lai:2004,
Samoilov:2005, Elf:2003} have been formulated and analyzed
numerically with the aim of identifying the sources of randomness,
and how noise is controlled and harnessed in the cellular
environment.

The starting point for such stochastic descriptions is the Master
Equation (ME), a conservation equation that gives the time
evolution of the probability distribution of the state of the
system. For a discrete state space, it can be written
as~\citep{vanKampen:1992}:
\begin{equation}
  \label{eq:cme}
  \dot{P}_j(t) = \sum_i W_{ji}P_i(t) - W_{ij}P_j(t),
\end{equation}
where $P_j(t)$ is the probability of finding the system in state
$j$ at time $t$ and $W_{ji}$ is the transition rate from state $i$
to state $j$. The application of the ME to chemical systems goes
back at least to the study of an irreversible reaction
by~\citet{Delbruck:1940}. This work has been extended by a number
of authors (see~\citet{McQuarrie:1967} and~\citet{Paulsson:2005}
for reviews). Eq. \ref{eq:cme} is usually referred to as the
Chemical Master Equation (CME) when it describes stochastic
versions of Law of Mass Action systems. In this paper, we extend
the notation and apply the term CME to compound reaction rates
(e.g., Michaelis-Menten) as well.

Although theoretically rigorous, there are very few systems for
which the ME has been solved analytically. Direct numerical
integration is often very difficult due to the fact that the state
space grows very rapidly and also to the stiffness of the
equations. One way to overcome these issues is to employ some kind
of approximation scheme. In certain limits, one can consider
continuum approximations leading to partial differential
equations, such as van Kampen's Linear Noise Approximation and
Fokker-Planck equations~\citep{vanKampen:1992}. However, these
approximations disregard the discreteness of the state of the
system, and can therefore give rise to significant deviations when
the number of molecules is very small, as can be the case for
GRNs.

A different approach for dealing with the CME numerically is
provided by the widely used Stochastic Simulation Algorithm (SSA)
by~\citet{Gillespie:1976}. The SSA is a kinetic Monte Carlo
algorithm, rigorously derived from the same assumptions as the
CME, which gives realizations of the trajectory of a given CME.
The SSA follows an idea that can be traced back to
Doob~\cite{Doob:1945} and provides an \emph{exact} procedure for
the kinetic sampling of the CME~\citep{Gillespie:1992} in the
sense that we obtain an unbiased and convergent estimate of the
solution of the CME. Due to its biological applications, there has
been considerable interest in the SSA in recent years and several
extensions have been proposed (e.g., the explicit introduction of
space~\citep{Stundzia:1996, Elf:2004} and time-delays
\citep{Bratsun:2005}), as well as algorithmic speed-up
improvements~\citep{Gibson:2000, Cao:2006b}.

In many experiments of interest~\citep{Gardner:2000, Pedraza:2005,
Fung:2005, Becskei:2001}, and in related theoretical
analyses~\citep{Hasty:2000, Kepler:2001, Lai:2004, McAdams:1997,
Thattai:2001, Tomioka:2004, Kummer:2005, Elf:2003}, the system at
hand is assumed to have reached the stationary distribution, i.e.,
the left hand side of Eq.~\ref{eq:cme} is zero. Consider the
ME~(Eq.~\ref{eq:cme}) in operator form:
\begin{equation}
  \label{eq:me_eigen}
  \dot{\mathbf{P}} = W \mathbf{P} - \mathrm{diag}(\mathbf{e}^T W)\mathbf{P} \equiv Q^T\mathbf{P},
\end{equation}
where $\mathbf{P}$ is a (possibly infinite-dimensional) column vector
of probabilities, $\mathbf{e}$ is the vector of ones, and $W$ is the
transition matrix.  The stationary distribution $\mathbf{P}^*$
(usually denoted $\pi$ in Markov theory) could in principle be
obtained from the first left eigenvector of the
$Q$-matrix~\citep{Norris:1999}. However, this approach is impractical
due to the curse of dimensionality: the run-time and memory
requirements scale exponentially with the number of types of
molecules. For small state spaces, one can still consider a
(truncated) finite subspace to obtain an approximate convergent
solution. Our own accurate version of this \emph{approximate
eigenvector method} is used in this paper to check the accuracy of our
DCFTP-SSA algorithm when an analytical expression for the stationary
distribution is not known.

In order to sample from the stationary solution of
Eq.~\ref{eq:cme}, one would have to run the SSA for an `infinite'
time.  Of course, the most common practical solution is to run the
algorithm `repeatedly' for a `very long time' and hope that the
system has reached the stationary distribution when the run is
stopped~\citep{Erban:2006}. The lack of a termination certificate
can lead to computational inefficiency if the runs are longer than
necessary or, more importantly, to mis-sampling from the wrong
distribution if the runs are not long enough.

In this paper we present an algorithm (DCFTP-SSA) that guarantees
perfect sampling of the stationary solution of the CME for the general
class of biochemical networks formed by the interconnection of
generalized Hill- and Monod-type reactions, the canonical models for
GRNs and enzymatic networks. The DCFTP-SSA builds on the standard SSA
and considers it in the light of Markov Chain Monte Carlo (MCMC)
algorithms, specifically within the Dominated Coupling From The Past
(DCFTP) framework introduced by~\citet{Propp:1996}
and~\citet{Kendall:1997}. Henceforth, we introduce the algorithm and
its properties in detail through the analysis of the building blocks
of GRNs.  We then apply the DCFTP-SSA to the study of two systems of
experimental interest in synthetic biology: the toggle
switch~\citep{Gardner:2000} and the
repressilator~\citep{Elowitz:2000}.

\section*{The Dominated Coupling From The Past --\\ Stochastic Simulation Algorithm (DCFTP-SSA)}

In many applications we are interested in sampling from a complicated
distribution that cannot be written down explicitly.  Markov Chain
Monte Carlo (MCMC) methods can sometimes provide an answer by setting
up a Markov chain which has the desired distribution as its stationary
distribution. Sampling from the target distribution is then achieved
by evolving the Markov chain until it has reached its stationary
distribution~\citep{Norris:1999}. The major issue in these schemes is
when to stop the Markov chain. This problem was addressed by Propp and
Wilson with their Coupling From The Past (CFTP)
algorithm~\citep{Propp:1996}, a celebrated example of what are
commonly referred to as \emph{Perfect Sampling}
algorithms~\citep{Thonnes:2000}. The version used here is the
Dominated Coupling From The Past (DCFTP) algorithm, the extension
introduced by Kendall~\citep{Kendall:1997} to study continuous-time
Markov processes on an unbounded state space. We now give a brief
introduction to the algorithm.

In principle, a Markov chain would have to be run for an infinite time
in order to reach its stationary distribution. The CFTP algorithm,
however, recasts the problem as a procedure in which the running time
becomes a random variable but a certificate is issued if the
stationary distribution has been reached. This is signalled by the
coalescence of relevant coupled Markov chains.  Markov chains are
\emph{coupled} if they have the same transition rules and use the same
sequence of random numbers for their realization. Coupled Markov
chains are said to \emph{coalesce} when they meet. Clearly, coupled
chains with the same transition rule (but started from different
initial states) will have the same values for all future times after
the coalescence time $T_c$. Propp and Wilson proposed to use Markov
chains \emph{coupled from the past} to ensure that the whole state
space of initial conditions maps to the same state at present. By
extending sufficiently far back into the past and using a fixed
sequence of random numbers, we will eventually reach a time from which
all paths map to the same state at $t = 0$.  This condition is
equivalent to the stationary distribution, since the starting
condition is irrelevant for the current state.

\begin{figure}
  \centering
  \includegraphics[width = 3.25in]{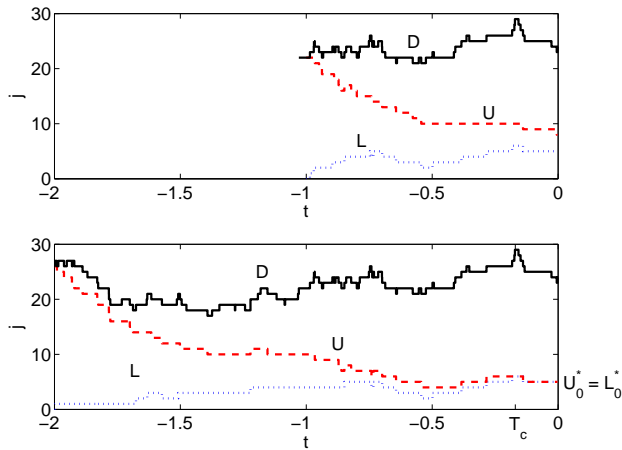}
  \caption{\textbf{An illustrative run of the DCFTP-SSA for the Hill
equation.} The top panel shows the algorithm started from $t =
-1$. Both the upper ($U$) and lower ($L$) are coupled to the
dominating path ($D$) and here they fail to coalesce before $t =
0$. The coupling implies that $U$ and $L$ can only change states if
$D$ does so. In the lower panel, the algorithm is restarted from $t =
-2$. Note that $D$ extends further into the past but the realization
of $D$ in the interval $t \in [-1, 0]$ is the same as above. In this
case, coalescence occurs at $T_c = -0.17$ and the value $U^*_0=L^*_0$
is guaranteed to be a sample from the stationary solution of the Hill
CME~(Eq.~\ref{eq:hill}).  }
  \label{fig:dcftp}
\end{figure}

In many systems, the state space is very large, making it
infeasible to monitor all paths and their coalescence.  However,
the situation is much simpler if the state space is
\emph{partially ordered} and the partial ordering is maintained by
the transition rules. This is the case if we have a
\emph{monotone} transition rule $\phi$, such that $\phi(x, R) \leq
\phi(y, R) \quad \mathrm{if} \ x \preceq y$ , where $\preceq$
denotes the partial ordering, $x$ and $y$ are states and $R$ is a
random number used to determine the transition. The algorithm also
works for \emph{anti-monotone} transition rules: $\phi(x, R) \geq
\phi(y, R)$ if $x \preceq y$~\citep{Haggstrom:1998}. For a state
space to be partially ordered, it must be reflexive, antisymmetric
and transitive. If the monotonicity and partial ordering hold, we
only need to monitor two paths: an upper path $U$ and a lower path
$L$, since all other paths lie in between and will coalesce when
$U$ and $L$ have done so. This `sandwiching' property is
illustrated in Fig.~\ref{fig:dcftp}.  The guaranteed sampling of
the CFTP scheme comes with a trade-off: the run time is unbounded,
since $T_c$ is itself a random variable, but it is finite almost
surely. If the run is prematurely terminated by an impatient user,
the sample will be biased.

If the system has an unbounded state space, we must instead employ
Kendall's Dominated CFTP algorithm~\citep{Kendall:1997}. The DCFTP
relies on finding a reversible dominating process $D$ which bounds
the original process from above and for which the stationary
distribution is known. To create a dominating process, we exploit
the reversibility and stationarity of $D$: start the chain
$\tilde{D}$ at $t = 0$ from the stationary distribution and evolve
it until $t = T$; then use $D_{-t} = \tilde{D}_t$ as the
dominating process on the interval $[-T, 0]$.  It then follows
that \emph{all} chains of the original dominated process started
at $t = -\infty$ will be less than or equal to $D$ at time $-T$
and, consequently, chains coupled to $D$ started from a state
$U_{-T} \preceq D_{-T}$ can be interpreted as random realizations
of the original (dominated) process started from $t=-\infty$.

Gillespie's SSA implements a continuous time Markov process. As
such, it can be reformulated in the CFTP framework if the state
space is partially ordered and the propensity functions are
(anti-)monotone. Indeed, one can show that a partial ordering
based on the number of molecules in each species is maintained for
systems of unimolecular reactions with (anti-)monotone propensity
functions, like those generated by Hill or Monod birth rates
(Hemberg and Barahona, unpublished). If such a system has a fixed
(or bounded) number of molecules, the upper path is trivial and
the standard CFTP can be directly applied to the SSA.  However, in
most situations the number of molecules is unbounded and we must
use the DCFTP algorithm~\citep{Kendall:1997}.

Based on the discussion above, a perfect sampling DCFTP-SSA can be
developed if two pre-requisites are fulfilled. Firstly, the Markov
process defined by the dominating process must be reversible. This
means that for every reaction in the system, there must exist
another reaction with products and reactants exchanged. Note that
this is different to requiring that the underlying reactions be
reversible and it typically means that for every reaction creating
a molecule, there must exist a degradation reaction. Secondly, we
must find a dominating process for the particular CME with a known
stationary distribution. We will show below that this requirement
can be fulfilled for genetic and enzymatic networks formed by the
interconnection of Hill-type, Monod-type and linear enzymatic
unimolecular reactions. This is based on the fact that the CME of
networks of Hill-type or Monod-type elements can be bounded from
above by processes based on networks of first order reactions, for
which the stationary distribution is known to be the multivariate
Poisson~\citep{Gadgil:2005}.

A brief outline of the DCFTP-SSA is as follows:

\vspace*{.08in}
\framebox{
\begin{minipage}[c]{.8 \textwidth}
\vspace*{.02in}
\begin{small}
\begin{enumerate}
\item Run the dominating process $\tilde{D}$ with known stationary
  distribution forward in time from $t = 0$ until $t = T$.
\item Apply time-reversal to the stationary process $\tilde{D}$ to obtain
  the dominating process $D$ such that $D_t = \tilde{D}_{-t}$ from $t=-T$ to $t=0$.
\item Start upper ($U$) and lower ($L$) chains of the original process
starting from $U_{-T} = D_{-T}$ and $L_{-T}= 0$ and update each chain \textit{coupled} to
  $D$  until $t=0$.
\item If the chains have coalesced at $t =0$, the common value
$U_0^*=L_0^*$ is a sample from the target distribution $\mathbf{P^*}$.
\item If coalescence has not occurred, double the running time to
  $t=2T$ reusing the random numbers from the previous iteration and
  repeat.
\item Keep doubling the running time until coalescence has occurred at $t=0$.
\end{enumerate}
\end{small}
\vspace*{.02in}
\end{minipage}}
\vspace*{.08in}

The main feature of the DCFTP-SSA is that it provides a
certificate for correct sampling from the stationary distribution.
Therefore we can use the DCFTP-SSA to numerically sample the
stationary distribution of the system with guaranteed Monte Carlo
accuracy. This can be of importance in high-dimensional systems
with complicated landscapes where the SSA simulations can converge
slowly. In the first two sections below, we test the application
of the DCFTP-SSA to the sampling of stationary distributions. In
those cases we only use the final value of each run as a sample,
thus ensuring that the samples are independent.

In addition, we can use the algorithm to discard the `burn-in'
period in stochastic simulations by running the DCFTP-SSA until a
guaranteed sample from the stationary distribution is obtained and
then using it to initiate a run of the ordinary SSA from time
$t=0$ onwards. This is important for the numerical study of
stationary properties of systems with high variability, e.g., with
underlying oscillatory or excitatory behaviour. In the final
section, we present examples of these applications to the toggle
switch and the repressilator.

\subsection*{Detailed application to the first order reaction}

In order to illustrate the DCFTP-SSA, we study in detail the simple
one-dimensional first order reaction, for which a full
\textit{time-dependent analytical} solution of the CME is
available. This allows us to study the convergence of the algorithm,
as compared to the standard SSA, and to explore the difference between
two sources of error which often get entangled: finite sampling error
and the mis-sampling error due to the fact that we may not have
reached the true stationary distribution.

Consider the simple first order reaction, which has been used as a
very simplified description of transcription and
translation~\citep{Thattai:2001, Austin:2006}:
\begin{equation}
  \label{eq:first_reaction}
  \emptyset \ \xrightarrow{k} \ A \ \xrightarrow{1} \ \emptyset  \quad \Longrightarrow \quad \dot{A} = k - A,
\end{equation}
where species $A$ is created at a (normalized) constant rate $k$ from
a source ($\emptyset$) and degraded to a sink ($\emptyset$). The
corresponding CME is
\begin{equation}
  \label{eq:first_me}
  \dot{P}_j = kP_{j-1} - kP_j + (j + 1)P_{j+1} - jP_j \equiv (\Em - 1)kP_j + (\E - 1)jP_j,
\end{equation}
with $P_j$ denoting the probability of having $j$ molecules of $A$. \E\ and
\Em\ are the step operators defined by van Kampen
\citep{vanKampen:1992}: $\E f(j) = f(j + 1)$ and $\Em f(j) = f(j - 1)$
for a function $f(j)$.

\begin{figure}[t]
  \centering
  \includegraphics[width = 3.25in]{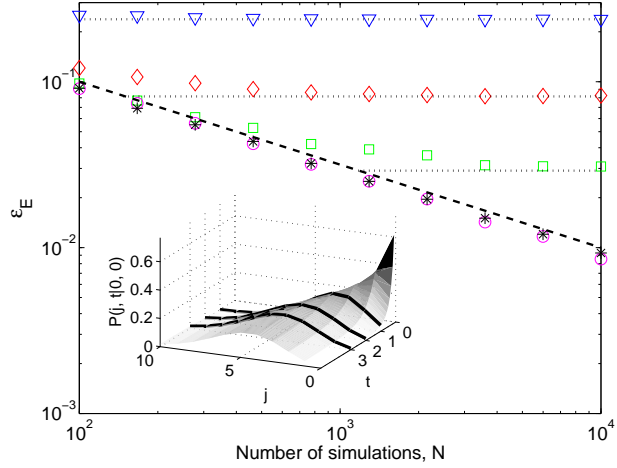}
  \caption{\textbf{Comparison of the convergence of the DCFTP-SSA vs.\ the
  standard SSA for the first order
  reaction~(Eq.~\ref{eq:first_me}) with $k = 5$.} We present the
  Euclidean error of the distribution as a function of the number of
  Monte Carlo runs $N$ for the DCFTP-SSA ($\ast$) and for the standard
  SSA with stopping times: $t_{\rm SSA} = 1 (\nabla), 2 (\Diamond), 3
  (\Box), 15 (\bigcirc)$. Note that the error of the SSA runs
  converges to the asymptotic error floors for the different stopping
  times given by Eq.~\ref{eq:first_euclidean_err} (dotted lines). The
  dashed line corresponds to the $N^{-1/2}$ Monte Carlo scaling and
  shows the correct convergence of the DCFTP-SSA (and the SSA only
  when the stopping time is with very long).  Each point is calculated
  from the mean of 100 different ensembles of $N$ samples, with
  numerical error bars smaller than the symbols.  \emph{Inset:} The
  full time-dependent solution of the first order
  reaction~(Eq.~\ref{eq:first_sol}) from a $\delta$-distribution
  initial condition.  The distribution evolves towards a smoother
  Poisson distribution. The dark lines correspond to the distributions
  at $t =1, 2, 3$ used in the main panel.}
  \label{fig:first_error}
\end{figure}

Equation~\ref{eq:first_me} is one of a few CMEs for which the analytical
expression of the stationary solution is known~\citep{Gadgil:2005,
vanKampen:1992}: it is the Poisson distribution with parameter
$\lambda = k$. More importantly for our purposes, one can obtain the
full \emph{time-dependent solution} of the CME (see
Appendix~\ref{app:first}). For the usual initial condition with $0$
molecules, the time-dependent distribution $\mathbf{P}(t)$ is given by
(see Fig.~\ref{fig:first_error} inset):
\begin{equation}
  \label{eq:first_sol}
  P_j(t) \equiv P(j, t \vert 0, 0) = \frac{\exp \left( -k(1 - e^{-t}) \right)}{j!} \left[ k(1 - e^{-t}) \right]^j  ,
\end{equation}
which converges to the correct stationary distribution as $t \to \infty$.

The mis-sampling error can be understood analytically in this
simple example.  If the standard SSA is used to estimate the
stationary solution of the CME~(Eq.~\ref{eq:first_me}), $N$
samples from independent SSA runs will be collected at a stopping
time $t_{\rm SSA}$.  This leads to the sampled distribution
$\mathbf{P}_{\rm SSA}(N,t_{\rm SSA})$, which clearly converges to
the stationary distribution $\mathbf{P}^*$ in the double limit
$t_{\rm SSA}, N \to \infty$.  As $N$ is increased, with fixed
$t_{\rm SSA}$, we sample with increasing Monte Carlo accuracy from
$\mathbf{P}(t_{\rm SSA})$ given by Eq.~\ref{eq:first_sol} but not
from the true stationary distribution $\mathbf{P}^*$.  We
therefore reach an error floor that cannot be broken.  Using the
analytical expression~(Eq.~\ref{eq:first_sol}), the asymptotic
error floor $\eucerr^*(t_{\rm SSA})$ is shown to be
\begin{eqnarray}
 \label{eq:first_euclidean_err}
\eucerr^*(t_{\rm SSA})^2 &=& \sum_{j = 0}^{\infty} \left(P_j^* - P_j(t_{\rm SSA}) \right)^2
  = \sum_{j = 0}^{\infty} \left( \frac{e^{-k}k^j}{j!} - \frac{e^{-\alpha k}(\alpha k)^j}{j!} \right)^2 \nonumber \\
  & =& I_0(2k)e^{-2k} - 2I_0(2k\sqrt{\alpha})e^{-k - \alpha} + I_0(2k\alpha)e^{-2\alpha},
\end{eqnarray}
where $\alpha = 1 - e^{-t_{\rm SSA}}$ and $I_0(x)$ is the modified
Bessel function of the first kind.
Fig.~\ref{fig:first_error} shows that the levelling-off of the Euclidean error for
$\mathbf{P}_{\rm SSA}(N,t_{\rm SSA})$ is explained by the
error floors calculated from Eq.~\ref{eq:first_euclidean_err}.

This source of error is eliminated in a DCFTP-SSA formulation of this
process.  The key step is to find a reversible dominating process with
known stationary distribution. In this example, it is clear that the
best choice is to use the process~(Eq.~\ref{eq:first_me}) itself,
i.e., the $U$ and $D$ chains in Fig.~\ref{fig:dcftp} will be
identical. Figure~\ref{fig:first_error} shows the Euclidean error for
the DCFTP-SSA sample distribution, which shows no flooring and the
expected $N^{-1/2}$ scaling with the number of Monte Carlo
runs~\citep{Shreider:1966}. Equivalent results are obtained with other
error measures for the sampled distribution, e.g., the
Kullback-Leibler and Kolmogorov distances, and the $\chi^2$-goodness
of fit.

\section*{The building blocks: networks and nonlinear elements}
In order to extend our study to GRNs, we need to consider how to deal with two key ingredients:
networks of reactions, and nonlinear elements of the Hill and Monod types.

\subsection*{Application to networks: Simple gene model of two coupled first order reactions}

To showcase how to apply the DCFTP-SSA to networks of several species,
we use a widely studied, simplified model of gene
expression~\citep{Kaern:2005, Thattai:2001, Paulsson:2004}.  In its
simplest form, the model consists of two types of molecules: mRNA
($M$) and proteins ($R$), which are produced and depleted at constant
rates. In addition, the mRNA catalyzes protein production through a
linear enzymatic reaction:
\begin{center}
  \begin{pspicture}(0, 2.75)(3, 5)
    \rput(0, 4){$\emptyset$}
    \psline[linewidth=.5pt]{->}(0.2, 4)(1, 4)
    \rput(0.55, 4.25){\small{$k_M$}}
    \rput(1.2, 4){$M$}
    \psline[linewidth=.5pt]{->}(1.2, 3.8)(0.7, 3.4)
    \rput(0.75, 3.78){\small{$k_R$}}
    \psline[linewidth=.5pt]{->}(1.5, 4)(2.3, 4)
    \rput(1.9, 4.2){\small{1}}
    \rput(2.5, 4){$\emptyset$}
    \rput(0, 3.3){$\emptyset$}
    \psline[linewidth=.5pt]{->}(0.2, 3.3)(1, 3.3)
    \rput(0.55, 3.12){\small{$k_B$}}
    \rput(1.2, 3.3){$R$}
    \psline[linewidth=.5pt]{->}(1.5, 3.3)(2.3, 3.3)
    \rput(1.9, 3.1){\small{1}}
    \rput(2.5, 3.3){$\emptyset$}
 \end{pspicture}
\end{center}

\begin{figure}[t]
  \centering
  \includegraphics[width = 3.25in]{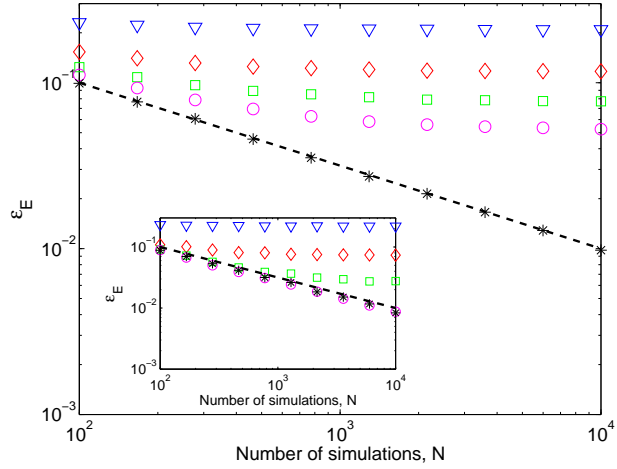}
  \caption{\textbf{The effect of dimensionality on the convergence of
  the sampled distribution. The main panel shows the Euclidean error
  for the joint distribution of the linear gene
  model~(Eq.~\ref{eq:gene_me}) with $\kM = 3$, $\kR = 3$ and $\kB =
  1$.}  The open symbols correspond to the standard SSA with stopping
  times as in Fig.~\ref{fig:first_error} while the DCFTP-SSA is marked
  by asterisks. The points are obtained by averaging over 100
  ensembles of $N$ samples, with error bars contained within the
  symbols. The higher dimensionality and the topology of the network
  make the convergence of the standard SSA much slower for this case
  than for the one-dimensional first order reaction in
  Fig.~\ref{fig:first_error}. Note that the DCFTP-SSA is unaffected by
  error floors and shows $N^{-1/2}$ convergence, as given by the
  dashed line. \emph{Inset:} Same as the main panel for the marginal
  distribution of the mRNA ($M$).  In this case, the convergence is as
  for the one-dimensional linear reaction in
  Fig.~\ref{fig:first_error}.  The difference between the convergence
  of the joint and marginal distributions is not surprising: the
  production of $R$ is downstream from $M$ and it is unlikely to reach
  the stationary level before $M$.}
  \label{fig:gene_error}
\end{figure}

The deterministic description is given by
\begin{eqnarray}
  \dot{M}& = & k_M - M \nonumber \\
  \dot{R} & = & k_B + k_R M - R,
\end{eqnarray}
and the corresponding CME is
\begin{eqnarray}
\label{eq:gene_me}
\dot{P}_{M, R} & = & (\Epm{M} - 1)\kM P_{M, R} + (\Ep{M} - 1)MP_{M, R} \nonumber\\
& + & (\Epm{R} - 1)(\kB + \kR M)P_{M, R} + (\Ep{R} - 1)RP_{M, R}.
\end{eqnarray}
In fact, Eq.~\ref{eq:gene_me} has been solved: the stationary
state of the CME of any general network of linear reactions is a
multivariate Poisson distribution~\citep{Paulsson:2004, Gadgil:2005}.
This means that, similarly to the one-dimensional first order
reaction, the DCFTP-SSA is directly applicable to
Eq.~\ref{eq:gene_me}, since the reaction network is reversible and
the multivariate Poisson stationary solution is itself a dominating
process for the system.  The existence of an analytical solution allows us
to check how the accuracy of the DCFTP-SSA depends
on the dimensionality of the system. Fig.~\ref{fig:gene_error}
(inset) shows that the marginal distribution (for $M$) converges
like the one dimensional reaction in Fig.~\ref{fig:first_error}, with
similar error floors for the standard SSA. Note, however, that the joint
distribution (Fig.~\ref{fig:gene_error}) exhibits slower convergence of
the standard SSA due to the higher dimensionality of the system,
indicating the need for longer SSA runs to avoid mis-sampling from the
true distribution. The DCFTP-SSA shows $N^{-1/2}$ Monte Carlo
scaling with no error floors regardless of the dimensionality of the
system.

\subsection*{Non-linear elements: Hill and Monod reactions}

The \emph{Hill reaction} is widely used to model the sigmoidal
(non-linear) characteristics of many biological processes and
specifically those involved in genetic regulation~\citep{Gardner:2000,
Elowitz:2000, Cornish-Bowden:2004}. This model incorporates negative
feedback, whereby the rate of creation of new molecules decreases as
their concentration increases:
\begin{equation}
  \label{eq:hill_reac}
  \emptyset \xrightarrow{f(A; \, k, \alpha)} A \xrightarrow{1} \emptyset \quad \Longrightarrow \quad \dot{A} = \frac{k}{\theta^\alpha + A^{\alpha}} - A,
\end{equation}
where $f(A; k, \alpha) = k/(\theta^\alpha + A^{\alpha})$ is the
state-dependent reaction rate, $k$ is the renormalized reaction
constant and $\alpha$ is the cooperativity
factor~\citep{Rosenfeld:2002}. Eq.~\ref{eq:hill_reac} can be
derived from elementary Law of Mass Action kinetics through
elimination of variables~\citep{Kepler:2001}.  The corresponding CME is
given by:
\begin{equation}
  \label{eq:hill}
  \dot{P}_j = (\Em - 1)\frac{k}{\theta^\alpha + j^{\alpha}}P_j + (\E - 1)jP_j.
\end{equation}
It is usual to fix $\theta =1$ and we do so in the following.  For the
particular case $\alpha =1$, Eq.~\ref{eq:hill_reac} reduces to the
familiar Michaelis-Menten equation, for which we can obtain the
following analytical expression for the stationary distribution (see
Appendix~\ref{app:hill_monod}): $P^*_j = ck^j/(j!)^2$, where $c =
1/I_0(2\sqrt{k})$ is a normalization constant.

The \emph{Monod reaction} is used to model gene upregulation, as
referred in particular to the auto-catalysis common in many
eucaryotes~\citep{Becskei:2001, Cornish-Bowden:2004, Xiong:2003}. The
reaction can be written as
\begin{equation}
  \label{eq:monod_reac}
  \emptyset \xrightarrow{g(A; \, k, \alpha)} A \xrightarrow{1} \emptyset \quad \Longrightarrow \quad \dot{A} = \frac{kA^{\alpha}}{\theta^{\alpha} + A^{\alpha}} - A,
\end{equation}
where $g(A; k, \alpha) = k A^{\alpha}/(\theta^{\alpha} +
A^{\alpha})$ is the state-dependent reaction rate that encapsulates
the positive feedback. The corresponding CME is given by
\begin{equation}
  \label{eq:monod}
  \dot{P}_j = (\Em - 1)\frac{kj^{\alpha}}{\theta^{\alpha} + j^{\alpha}}P_j + (\E - 1)jP_j.
\end{equation}
Again, we fix $\theta =1$ in the following. The stationary
distribution of the Monod CME~(Eq.~\ref{eq:monod}) with $\alpha =1$
can be obtained analytically (see Appendix~\ref{app:hill_monod}):
$P^*_j = ck^j/j!$, with normalization constant $c = 1/(e^k + k - 1)$.

For general $\alpha$, no explicit solution for the stationary
distribution of the CMEs Eq.~\ref{eq:hill} and Eq.~\ref{eq:monod}
is known, and numerical methods are therefore necessary.
Fortunately, the DCFTP-SSA is directly applicable to the
reversible Hill and Monod CMEs: the first order reaction, studied
in detail in the previous Section, can be used to produce a
dominating process for both processes. To see this, note that
Eq.~\ref{eq:hill} and Eq.~\ref{eq:monod} have birth rates that are
bounded from above by $k$ at all times. It is thus straightforward
to make sure that the upper and lower paths have rates that
fulfill the requirements of the algorithm. As an illustration,
Fig.~\ref{fig:dcftp} shows one DCFTP-SSA run for the Hill
CME~(Eq.~\ref{eq:hill}), where $D$ corresponds to the dominating
first order linear process~(Eq.~\ref{eq:first_me}) and $U$ and $L$
are the coupled Hill processes~\citep{Thonnes:2000, Kendall:1997}.
Our detailed simulations of the Hill and Monod CMEs (data not
shown) show $N^{-1/2}$ convergence of the DCFTP-SSA unaffected by
the mis-sampling errors that can appear when using the standard
SSA with short stopping times.

\section*{Application to nonlinear models of Gene Regulatory Networks}

Our foregoing discussion clarifies why the DCFTP-SSA is applicable to
reversible GRNs consisting of interconnected linear enzymatic, Hill
and Monod birth reactions: it is possible to define a dominating
process for these networks based on the associated linear network for
which the stationary solution (a multivariate Poisson distribution) is
known.  We now apply our scheme to two recent, canonical examples of
synthetic GRNs: a bistable genetic toggle switch~\citep{Gardner:2000},
with a bimodal stationary distribution; and the
repressilator~\citep{Elowitz:2000}, a simple synthetic oscillator.

\subsection*{Toggle switch: two coupled Hill equations}

An important characteristic of biochemical networks is the possibility
of multistability~\citep{Gardner:2000, Acar:2005, Hasty:2000,
Kepler:2001, Thattai:2001, Erban:2006, Craciun:2006, Xiong:2003}, as
exemplified by the toggle switch designed by Gardner \textit{et al} by
combining two mutually repressing genes~\citep{Gardner:2000,
Kepler:2001, Erban:2006, McAdams:1997, Hasty:2000}. This leads to a
bistable system with a sharp switching threshold:
\begin{center}
  \begin{pspicture}(1, 2.5)(1, 5)
    \rput(-0.4, 4){$\emptyset$}
    \psline[linewidth=.5pt]{->}(-0.2, 4)(1, 4)
    \rput(0.4, 4.3){\scriptsize{$f(B; \, \kA, \alpha)$}}
    \rput(1.2, 4){$A$}
    \psline[linewidth=.5pt]{-|}(1.05, 3.85)(0.4, 3.4)
    \psline[linewidth=.5pt]{->}(1.4, 4)(2.3, 4)
    \rput(1.9, 4.3){\small{1}}
    \rput(2.5, 4){$\emptyset$}
    \rput(-0.4, 3.3){$\emptyset$}
    \psline[linewidth=.5pt]{->}(-0.2, 3.3)(1, 3.3)
    \rput(0.4, 3.05){\scriptsize{$f(A; \,  \kB, \beta)$}}
    \rput(1.2, 3.3){$B$}
    \psline[linewidth=.5pt]{-|}(1.05, 3.45)(0.4, 3.9)
    \psline[linewidth=.5pt]{->}(1.4, 3.3)(2.3, 3.3)
    \rput(1.9, 3.1){\small{1}}
    \rput(2.5, 3.3){$\emptyset$}
  \end{pspicture}
\end{center}
 A deterministic model can be derived using two coupled Hill-equations:
 \begin{eqnarray}
  \label{eq:toggle_det}
  \dot{A} & = & \frac{k_A}{1 + B^{\alpha}} - A \nonumber\\
  \dot{B} & = & \frac{k_B}{1 + A^{\beta}} - B,
\end{eqnarray}
where $A$ and $B$ are the molecular numbers of the transcription
factors. With appropriate cooperativity factors $\alpha, \beta >
1$ and reaction constants \kA\ and \kB, Eq.~\ref{eq:toggle_det}
has two stable fixed points~\citep{Gardner:2000}.

\begin{figure}[t]
  \begin{center}
    \includegraphics[width = 3.25in]{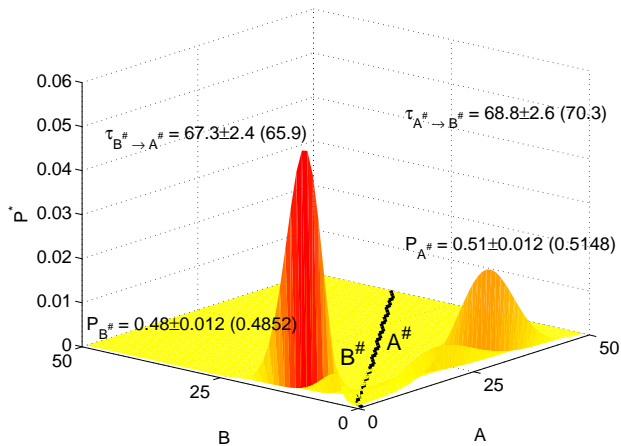}
  \end{center}
  \caption{\textbf{Stationary distribution for the toggle
    switch~(Eq.~\ref{eq:toggle}) with parameters $k_A = 30, k_B = 10,
    \alpha = 3, \beta = 1$.} The bimodality of the distribution leads
    to switching behavior. We have checked that the Euclidean error of
    the DCFTP-SSA sampled distribution decreases at the expected
    $N^{-1/2}$ rate (data not shown). The state space can be divided
    in two regions \abasin\ and \bbasin\, corresponding to the two
    basins of attraction of the fixed points.  The separatrix is found
    from the sign of the Fiedler eigenvector of the
    Laplacian~\citep{Chung:1997} of the state space lattice. The
    probabilities $P_{\abasin}$ and $P_{\bbasin}$ of finding the
    system in each basin computed with the DCFTP-SSA compare very well
    with those obtained with the approximate eigenvector method (in
    parenthesis).  Similarly, the expected time to reach the
    separatrix averaged over all the states in each
    basin~\citep{Norris:1999} calculated with the DCFTP-SSA (averaged
    over 25 ensembles, each consisting of 1000 samples) compare well
    with the approximate eigenvector method (in parenthesis).}
  \label{fig:toggle}
\end{figure}

The corresponding CME is given by
\begin{eqnarray}
  \label{eq:toggle}
  \dot{P}_{A, B} & = & (\Em_A - 1)\frac{k_A}{1 + B^{\alpha}}P_{A, B} + (\E_A - 1)AP_{A, B} \nonumber\\
 & + & (\Em_B - 1)\frac{k_B}{1 + A^{\beta}}P_{A, B} + (\E_B - 1)BP_{A, B}.
\end{eqnarray}
Based on our discussion above, it is easy to see that linear
process with constant birth rates \kA\ and \kB\ will be a
dominating process for Eq.~\ref{eq:toggle}. Since the
Hill-function is anti-monotone, we must use a \emph{cross-over}
scheme for the updates to make sure that all initial conditions
are sandwiched between the upper and lower paths, as described
in~\citet{Haggstrom:1998}. We have applied the DCFTP-SSA to
Eq.~\ref{eq:toggle} to obtain the stationary distribution of the
system, which is clearly bimodal (Fig.~\ref{fig:toggle}). Because
there is no analytical solution in this case, we have checked the
accuracy and convergence of the DCFTP-SSA against the approximate
eigenvector method with excellent results (not shown).

Beyond certifying proper convergence, the DCFTP-SSA can be used to
study properties of the network that need to be properly averaged over
the stationary distribution. Consider the mean time for the system to
switch from the basin of one fixed point \abasin\ to the other
\bbasin, i.e., the escape time of the CME~(Eq.~\ref{eq:toggle}).  If
the standard SSA were to be used, it would be unclear how to collect
proper Monte Carlo statistics, since no certificate of stationarity
exists.  In contrast, we use a mixed scheme in which an initial
DCFTP-SSA run is followed by a (faster) SSA run.  The initial
DCFTP-SSA run ensures that the \emph{initial condition} for the SSA
run is correctly sampled from the stationary distribution at $t=0$, in
essence eliminating the `burn-in' period. From that point onwards, a
standard SSA run is performed to obtain statistics of the average time
to cross the separatrix. The results are reported in
Fig.~\ref{fig:toggle}.

In their experiments, ~\citet{Gardner:2000} were able to induce
faster $\abasin \to \bbasin$ switching by reducing the
downregulation capability of $A$ in the cellular environment. This
can be mimicked by a modified CME in which the $A^{\beta}$-term in
Eq.~\ref{eq:toggle} is removed. If we run the SSA of this modified
CME starting from the stationary distribution of the original
system~(Eq.~\ref{eq:toggle}), the escape time $\tau_{\abasin \to
\bbasin}$ becomes approximately two orders of magnitude smaller,
in broad agreement with the experiment. In fact, it can be shown
that the stationary distribution of the modified CME becomes
unimodal, with no observable peak for species $A$.

\subsection*{Repressilator: six-dimensional system of linear enzymatic and Hill equations}

\begin{figure}
  \begin{center}
    \includegraphics[width = 2.5in]{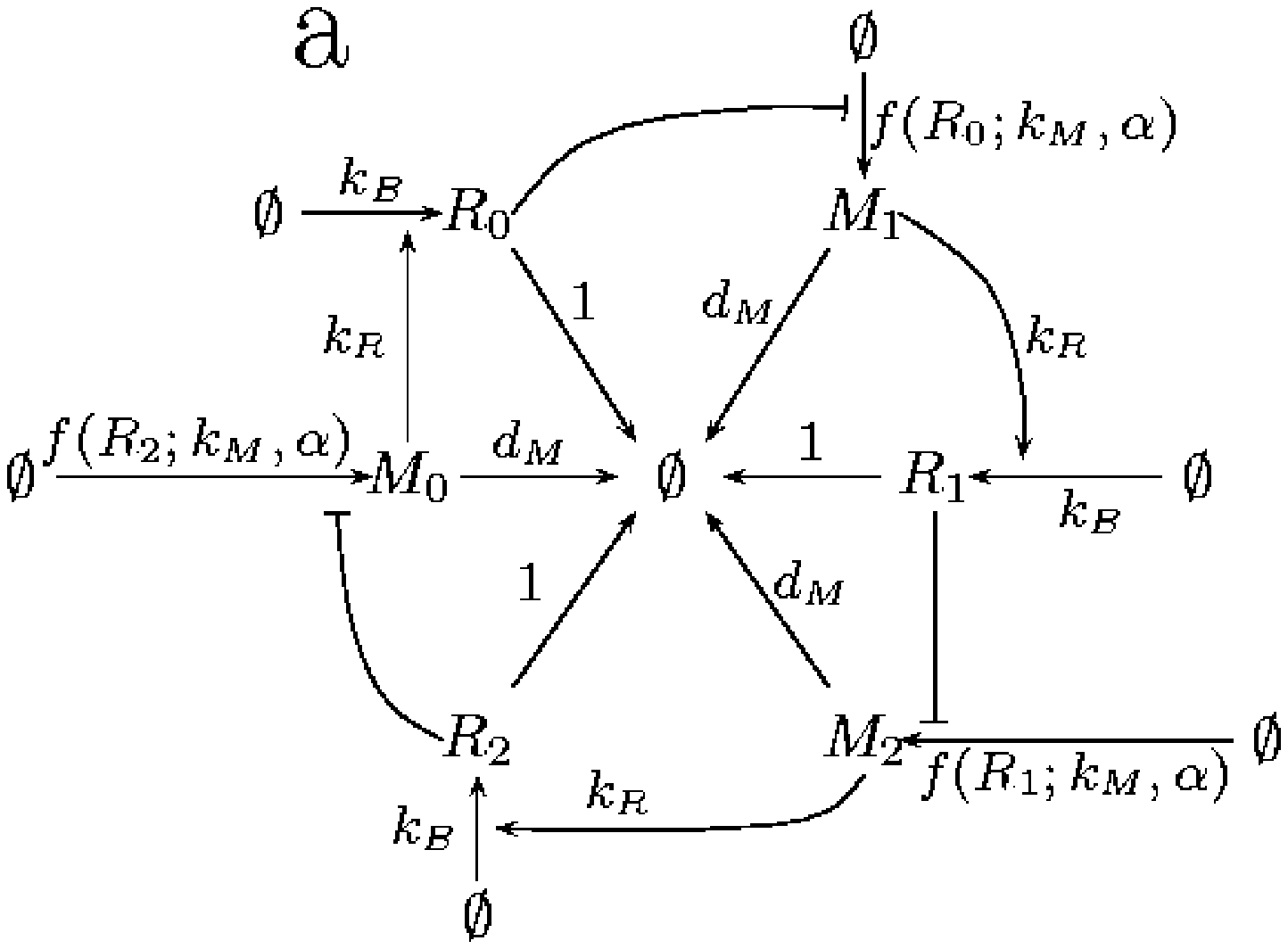}
    \includegraphics[width = 2.5in]{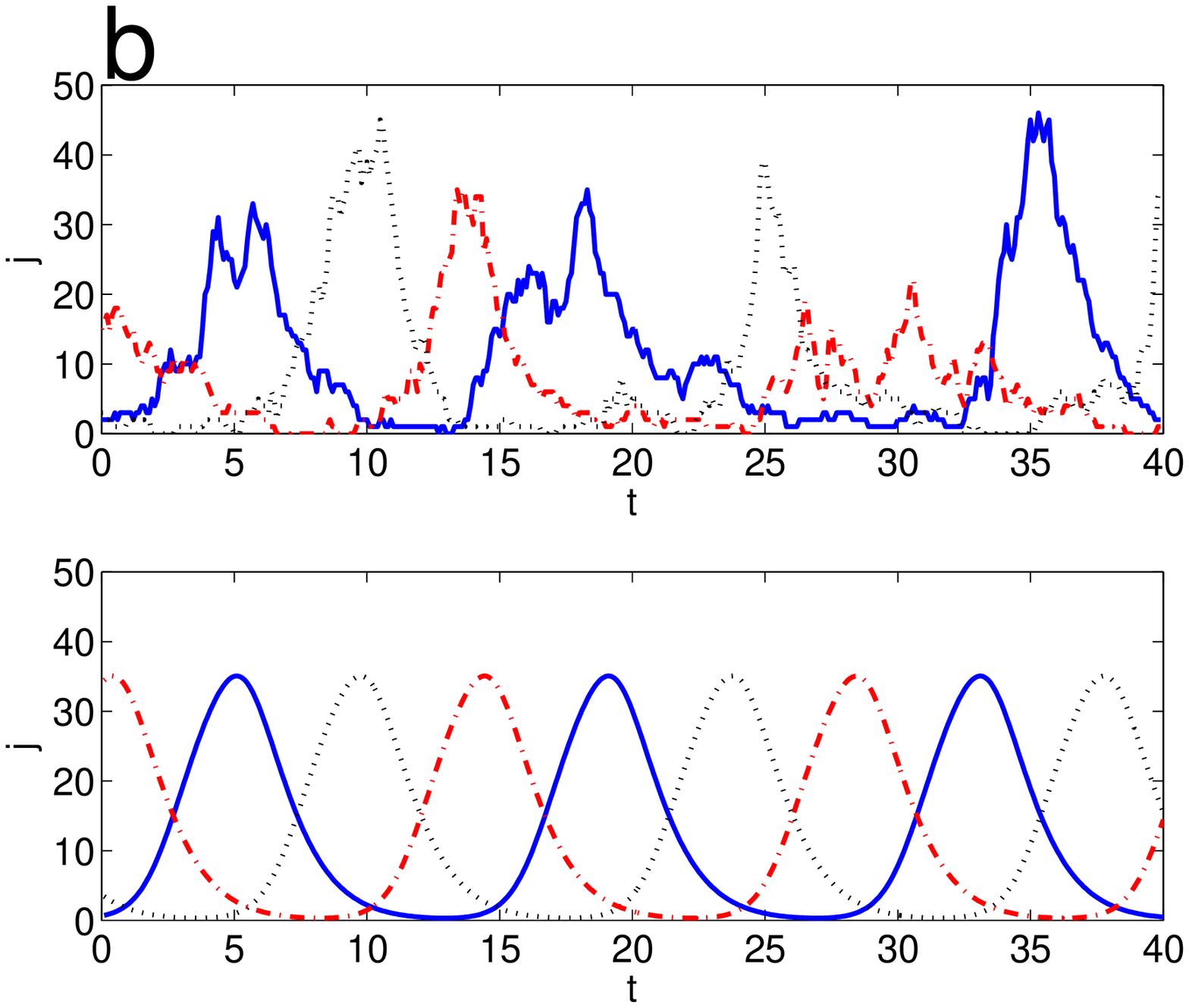}\\
    \includegraphics[width = 2.5in]{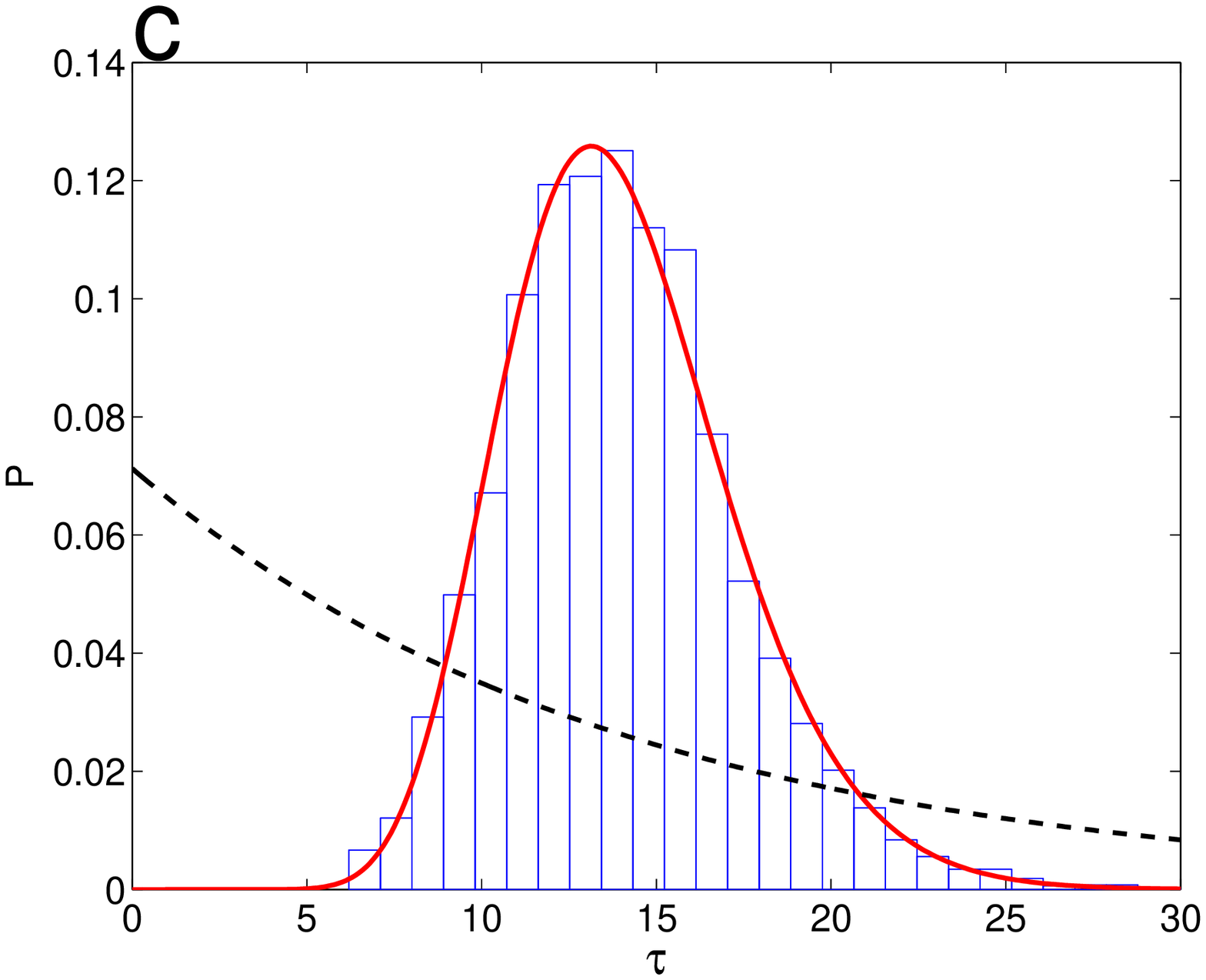}
    \includegraphics[width = 2.5in]{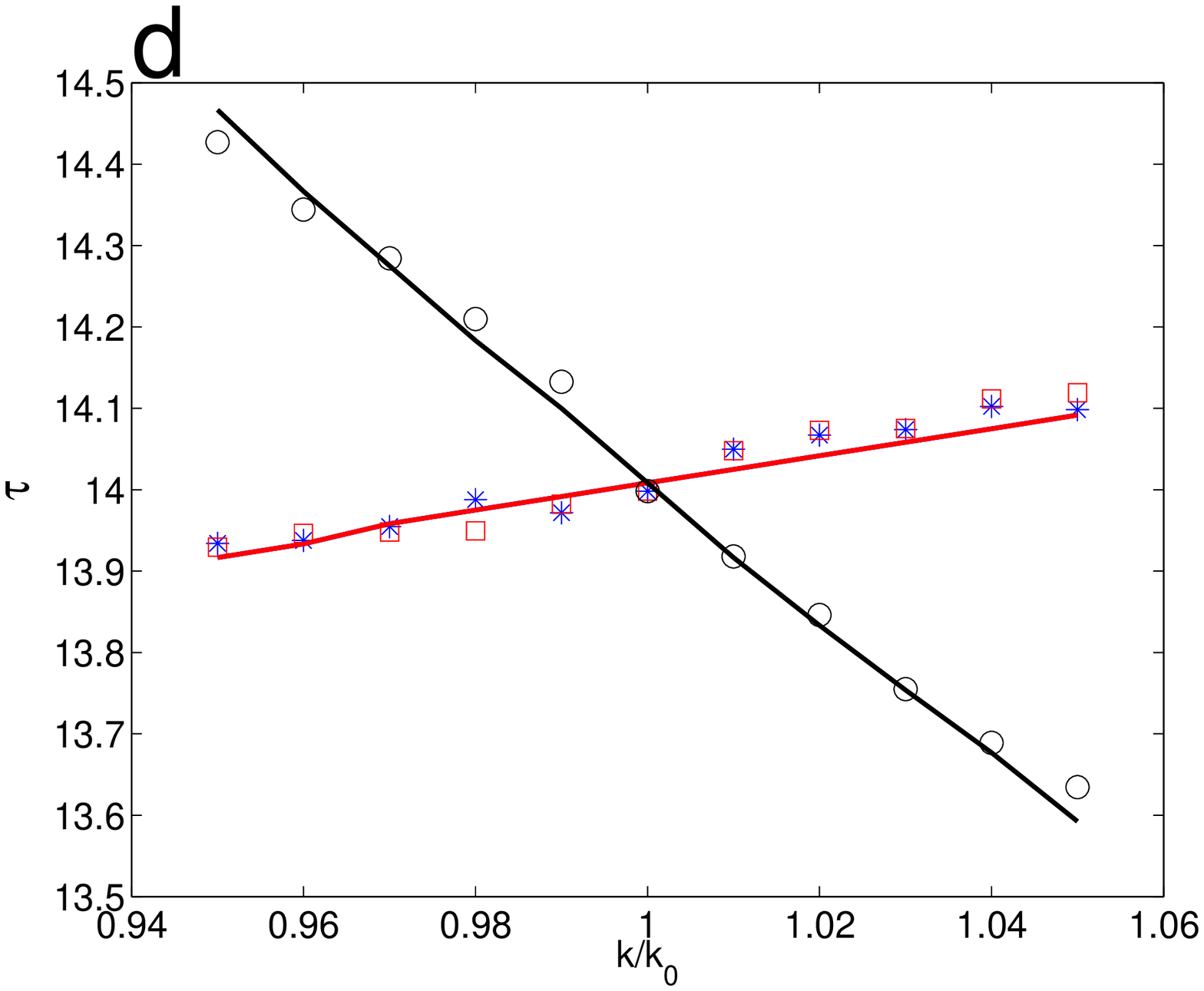}
    \caption{\textbf{Analysis of the repressilator with $\kM = 10$,
	$d_M = 1$, $\alpha = 2$, $\kB = 1$, $\kR = 3$.} (\textit{a})
      Diagrammatic representation of the repressilator network showing the
      alternating positive and negative feedbacks. (\textit{b})
      \textit{Top:} One SSA realization of the
      CME~(Eq.~\ref{eq:repressilator_cme}) started from the stationary
      distribution showing the concentrations of the three proteins. The
      oscillations are very noisy and far from sinusoidal.  The Fourier
      spectrum shows no distinguishable peaks (data not
      shown). \textit{Bottom:} The corresponding deterministic
      solution~(Eq.~\ref{eq:repressilator}) exhibits regular oscillations
      of the proteins in succession.  (\textit{c}) Distribution of the
      period $\tau$ at stationarity obtained from a long SSA run. The
      solid line corresponds to a best fit to a generalized Gamma
      distribution while the dotted line shows a Poisson distribution with
      the same mean.  (\textit{d}) Change of the period of the
      repressilator as a function of parameter variation. There is good
      agreement between the sensitivity of the stochastic system with
      respect to a $\pm 5 \%$ change in $\kM$ ($\ast$), $k_P$ ($\Box$),
      and $d_M$ ($\bigcirc$), and the deterministic system (solid lines).
      The effect of $d_M$ on the duration of the period $\tau$ is greater
      and opposite to that of \kM\ and $k_P$, which are similar.}
    \label{fig:repressilator}
  \end{center}  
\end{figure}

The repressilator, a synthetic system of transcriptional regulators,
is perhaps the simplest biochemical oscillator that has been
implemented experimentally~\citep{Elowitz:2000}. It consists of three
genes in a loop, in which the expression of one gene is inhibited by
the product of another gene in succession
(Fig.~\ref{fig:repressilator}\textit{a}):
\begin{eqnarray}
\label{eq:repressilator}
\dot{M}_i &=& \frac{\kM}{1 + R_{i + 1}^{\alpha}} - d_M M_i \nonumber \\
\dot{R}_i &=& \kB + \kR M_i  - R_i, \quad \quad i=0,1,2  \;\;( \mathrm{mod} \,\, 3).
\end{eqnarray}
Here $M_i$ are the mRNA levels (with production rate \kM\ and
degradation rate $d_M$) and $R_i$ are the corresponding proteins (with
basal rate \kB\ and linear production rate \kR).  This model of the
repressilator is therefore a network of linear enzymatic
elements~(Eq.~\ref{eq:gene_me}) for the protein production terms and
Hill reactions~(Eq.~\ref{eq:hill}) to modulate the production of mRNA.
The deterministic system~(Eq.~\ref{eq:repressilator}) has been shown
to be oscillatory, with the concentrations of the three proteins
peaking in succession (Fig.~\ref{fig:repressilator}\textit{b}, top panel).

The corresponding CME is given by
\begin{eqnarray}
  \label{eq:repressilator_cme}
  \dot{P}_n & = & \sum_{i = 0}^2 (\Em_{M_i} - 1)\frac{\kM}{(1 + R_{i + 1}^{\alpha})}P_n + (\E_{M_i} - 1)d_M M_i P_n \nonumber\\
 & + & (\Em_{R_i} - 1)(\kB +\kR M_i) P_n + (\E_{R_i} - 1)R_i P_n,
\end{eqnarray}
where the shorthand $P_n$ denotes the state $P_{M_0, M_1, M_2, R_0,
R_1, R_2}$ and $i \,\, \mathrm{mod} \,\, 3$.  We have simulated this
six dimensional system using the cross-over scheme described above: we
use the DCFTP-SSA to discard the `burn-in period', i.e., the time it
takes for the Markov process to reach the stationary distribution,
such that the initial conditions for the SSA runs are sampled from the
stationary distribution.  As a measure of the stationarity of our
DCFTP-SSA initialized sampling, we have checked its ergodicity by
establishing through an F-test that the average period and amplitude
of several short simulations are statistically indistiguishible from
those obtained from one long simulation.

As can be seen in the sample trajectory in
Fig.~\ref{fig:repressilator}\textit{b}, the stochastic system is
extremely noisy but it retains the overall oscillation of the
genes in succession. The oscillations do not die out even for very
long simulations and we have collected statistics of such long
runs. Fig.~\ref{fig:repressilator}\textit{c} shows a robust
feature of the oscillator: the distribution of the period is well
fitted by a generalized Gamma distribution, which is
characteristic of excitatory systems with a refractory period. The
oscillatory behaviour is present for a wide range of the
parameters $k_M, k_R, d_M$ \citep{Scott:2006}, which could be
modified experimentally. Fig.~\ref{fig:repressilator}\textit{d}
summarizes our investigation of the robustness of the
repressilator to parametric variation. Note the good agreement of
the calculated mean period with the corresponding deterministic
values.  The simulations show that changes in parameters $k_M$ and
$k_R$ produce a similar, small effect on the period.  The system
is, however, more sensitive to parametric changes in $d_M$. The
evaluation of how these parametric variations could be used for
the design of tunable and reliable oscillators will be the object
of further study~\citep{Tomshine:2006}.

\section*{Discussion}

In recent years, the importance of stochastic effects in gene
regulation has been elegantly demonstrated through experiments.
>From a theoretical perspective, such systems are usually analyzed
numerically with the standard SSA. When studying stationary
properties, this raises questions about the sampling of the
stochastic process since the SSA does not provide explicit
guarantees for convergence. One issue of interest for the use of
the DCFTP-SSA is that the shape of the stationary distribution can
be more important than the dimensionality of the system for the
convergence of the stochastic simulations. For systems with smooth
unimodal distributions, such as the repressilator, the SSA
approaches the stationary distribution rapidly, regardless of the
initial conditions. However, the ordinary SSA can be highly
sensitive to initial conditions for systems with multimodal
distributions, such as the toggle switch. In general, the SSA will
converge rapidly to the nearest mode. If we are primarily
interested in escape times, they can be estimated accurately by
using the ordinary SSA started close to one of the modes. However,
if the escape times from the modes are long and the ordinary SSA
is used, we run the risk of mis-sampling the stationary
distribution, specifically in important areas of low probability
such as the separatrix. The DCFTP-SSA circumvents this problem but
at the cost of longer coalescence times.

The guaranteed sampling provided by the DCFTP-SSA comes at an
extra computational cost.  In general, the coalescence time will
depend on the topology of the network and the form of the
reactions. However, the CPU overhead is in no way prohibitive and
the DCFTP-SSA can be run on an ordinary desktop computer for the
systems presented in this paper. An important theoretical feature
of the DCFTP-SSA is the fact that its runtime is not bounded. As
our simulations show, this feature does not seem to impose
practical restrictions on the algorithm. If necessary, however,
this issue can be resolved by using FMMR~\citep{Fill:2000},
another perfect sampling algorithm, which dispenses with variable
stopping times by running the Markov chains for a fixed time and
using rejection sampling to discard those simulations that have
not reached the stationary distribution.

In its current form, the DCFTP-SSA can be applied to reversible
systems of linear and nonlinear (anti-)monotonic unimolecular
reactions for which there exists a dominating process with known
stationary distribution. We note that the algorithm is applicable
not only to networks of linear enzymatic, Hill and Monod birth
reactions of standard form, but also to reactions whose rate
equation is a rational function that can be dominated by a linear
process. This condition is equivalent to requiring that the
polynomial in the denominator of the propensity function have a
higher degree than the polynomial in the numerator. These types of
rational functions are frequently obtained as compound
unimolecular rate laws to represent more complex
enzymatic~\citep{Cornish-Bowden:2004} or gene regulatory
reactions~\citep{Widder:2005}. An example of such functions is the
model for the $\lambda$-phage lysis-lysogeny
switch~\citep{McAdams:1997, Hasty:2000}, which we have also
simulated with the DCFTP-SSA (results not shown). Bimolecular and
two-substrate enzymatic reactions are not included in this group
and extending the DCFTP or CFTP to provide perfect sampling for
the generic SSA of second order reactions will entail further
research.

The DCFTP-SSA introduced here can be viewed as a reformulation and
extension of the Gillespie algorithm that provides guaranteed sampling
from stationarity for a generic class of systems relevant in GRNs,
thereby removing a source of uncontrolled uncertainty from the
simulations. By eliminating these extraneous sources of error,
comparisons between the predictions of different stochastic models at
stationarity can be performed more meaningfully. In the case of the
repressilator, we provided an accurate numerical characterization of
the stationary distribution for the period of this stochastic
oscillator. Because stationarity is guaranteed, the characteristics of
this distribution can be correlated with changes in the parameter
values (as shown in Figure \ref{fig:repressilator}\textit{d}) or
compared with those of other genetic oscillators. This could be a
useful tool for the analysis and design of synthetic transcriptional
oscillators and will be the subject of future research.

\vspace{1cm} \small{\textbf{Acknowledgments:} We would like to
thank Michael Elowitz for helpful comments about the
repressilator, and Matthieu Bultelle, Matias Rantalainen, Vincent
Rouilly and Sophia Yaliraki for valuable discussions and feedback.
We are also grateful for the constructive comments made by one of
the anonymous referees. Research funded by the EPSRC.}

\begin{appendix}

  \section{Time-dependent solution of the Master Equation for the one-dimensional first-order reaction} \label{app:first}

 The CME for the one-dimensional first order reaction
 (Eq.~\ref{eq:first_me}) has the same form as a homogenous birth-death process on the
  non-negative integers. It is also identical to the equation one obtains
  when studying the length of a $M/M/\infty$-queue with Poisson arrival
  and exponential service time~\citep{Norris:1999}.
  Associated with Eq.~\ref{eq:first_me}, there exists a
  family of orthogonal polynomials $\{\varphi_j(x)\}_{j=0}^{\infty}$ with
  the three term recurrence relation~\citep{Karlin:1955}:
  \begin{equation}
    \label{eq:birth-death_recurrence}
    -x \varphi_j(x) = j \varphi_{j - 1}(x) - (k + j)\varphi_j(x) + k \varphi_{j + 1}(x),
  \end{equation}
  where $\varphi_0(x) = 1$ and $\varphi_j(x) = 0$ for $j < 0$. Karlin
  and McGregor showed that the recurrence
  relation~(Eq.~\ref{eq:birth-death_recurrence}) leads to a spectral
  representation for $P_{ji}(t)$, the transition probability of going
  from state $i$ to state $j$ in time $t$:
  \begin{equation}
    \label{eq:first_spectral}
    P_{ji}(t) = P_j^* \int_0^{\infty} e^{-xt} \varphi_i(x)
    \varphi_j(x) \, d\phi(x),
  \end{equation}
  where $\phi(x)$ is a positive measure on $x$ and $P_j^*$ is the
  stationary distribution.

  The recurrence relation~(Eq.~\ref{eq:birth-death_recurrence}) is
  satisfied by the Charlier polynomials with $\phi(x)$ equal to the
  Poisson distribution~\citep{Karlin:1955}, i.e., the Charlier
  polynomials are a family of polynomials which are orthogonal with
  respect to the Poisson measure on a discrete lattice on the
  non-negative integers~\citep{Nikiforov:1991}. The Charlier polynomial
  of order $i$, parameter $a$ and argument $x$ is denoted
  \charlierlong{i}{a}{x}.

Using the following bilinear generating form of the Charlier
polynomials~\citep{Meixner:1939}
  \begin{equation}
    \label{eq:bilinear}
    \sum_{i=0}^{\infty} \frac{z^k}{k!} \charlierlong{i}{x}{m} \charlierlong{i}{y}{n} = e^z \left( 1 - \frac{z}{x} \right)^m \left( 1 - \frac{z}{y} \right)^n C_m \bigg( n; -\frac{(x - z)(y - z)}{z} \bigg),
  \end{equation}
  \citet{Lee:1997} showed that Eq.~\ref{eq:first_spectral} can be simplified to give:
  \begin{equation}
    \label{eq:first_transition}
    P_{ji}(t) = \frac{\exp \left( -k(1 - e^{-t}) \right)}{j!} \left[ k(1 - e^{-t}) \right]^j \left( 1 - e^{-t} \right)^i C_i \bigg(j; -\frac{k(1 - e^{-t})^2}{e^{-t}} \bigg).
  \end{equation}

The important special case with initial condition given by a
$\delta$-distribution at the origin can be simplified even further.
If $P_j(0) = \delta(0)$, then $C_0(j; a) = 1$ and
Eq.~\ref{eq:first_transition} becomes
  \begin{equation}
    \label{eq:first_sol0}
    P(j, t \vert 0, 0) = \frac{\exp \left( -k(1 - e^{-t}) \right)}{j!} \left[ k(1 - e^{-t}) \right]^j \equiv P_j(t),
  \end{equation}
  which is a Poisson distribution with time-varying parameter $k(t) =
  k(1 - e^{-t})$.  Therefore, the stationary Poisson distribution
  $P^*_j=k^j e^{-k}/j!$ is approached exponentially fast.

Note that
  \begin{equation}
    \lim_{t \rightarrow \infty} P_{ji}(t) = \frac{e^{-k}k^j}{j!},
  \end{equation}
which is another way of showing that the stationary distribution the process is Poisson no matter what the initial condition is.

  \section{Stationary solution of the Hill and Monod CMEs with $\alpha = 1$} \label{app:hill_monod}

 The stationary solution for the Hill CME (Eq.~\ref{eq:hill}) with
 cooperativity factor $\alpha=1$, which is equivalent to the
 Michaelis-Menten reaction, can be obtained analytically. This
 reaction can be viewed as a non-linear birth-death process, which
 allows us to write the stationary distribution $P_j^*$ as
 \citep{Norris:1999}
  \begin{equation}
    P_j^* = c\frac{\lambda_0 \lambda_1 \cdots \lambda_{j -1}}{\mu_1 \mu_2 \cdots \mu_j},
  \end{equation}
  where $\lambda_i$ and $\mu_i$ are the birth and death rates of state
  $i$ and $c$ is a normalization constant. Inserting the values from the
  CME~(Eq.~\ref{eq:hill}) we obtain
  \begin{equation}
    \label{eq:mm_sol}
    P_j^* = c^{-1}\frac{\frac{k}{\theta} \frac{k}{\theta+ 1} \cdots \frac{k}{\theta + j - 1}}{j!} =
    c^{-1}\frac{k^j\Gamma(\theta + 1)}{j!\Gamma(j + \theta)},
  \end{equation}
  with normalization constant
  \begin{displaymath}
    c = \frac{\theta I_\theta(2\sqrt{k}) + \sqrt{k}I_{1 + \theta}(2\sqrt{k})}{k^{\theta/2}},
  \end{displaymath}
  where $I_\theta(x)$ are Bessel functions.  For the special case of $\theta =
  1$, we use a recurrence relation for Bessel functions to simplify
  the above result to $c_0^{-1} = I_0(\sqrt{2}k)$.

 Consider now the Monod CME~(12) with $\alpha =1$:
  \begin{displaymath}
    \dot{P}_j = \left\{
      \begin{array}{ll}
    (\Em - 1)\frac{kj}{\theta + j}P_j + (\E - 1)jP_j & \textrm{if $j>1$} \\
    k & \textrm{if $j = 0$}.
      \end{array}
      \right .
  \end{displaymath}
  Note that the birth rate of the state $j = 0$ is modified to avoid
  it becoming absorbing. Using the same strategy as for the
  Michaelis-Menten CME above, one can find the stationary distribution
  \begin{equation}
    \label{eq:monod_sol}
    P_j^* = c\frac{k\frac{k}{\theta + 1} \cdots \frac{k(j - 1)}{\theta + j - 1}}{j!} =
    c\frac{k^j\Gamma(\theta + 1)}{j\Gamma(j + \theta)},
  \end{equation}
  with normalization constant $c = k\left(1+{}_1F_1(2, 1 + \theta, k)\right)$, where
  ${}_1F_1$ is Gauss' hypergeometric function.

\end{appendix}

\bibliographystyle{biophysj} \bibliography{virus}


\end{document}